\documentclass{jpp}

\usepackage{graphicx,bm}
\usepackage{natbib}
\newcommand{\bec}[1]{\mbox{\boldmath $ #1$}}

\title[Mean-field theory of differential rotation]{Mean-field theory of differential rotation in density stratified turbulent convection}

\author[I. Rogachevskii and N. Kleeorin]%
 {  I.\ns R\ls O\ls G\ls A\ls C\ls H\ls E\ls V\ls S\ls K\ls I\ls I
  \thanks{Email address for correspondence: gary@bgu.ac.il}
  \and
  N.\ns K\ls L\ls E\ls E\ls O\ls R\ls I\ls N
}
\affiliation{
Department of Mechanical Engineering, Ben-Gurion University of
the Negev, P. O. Box 653, 84105 Beer-Sheva, Israel
\\
Nordita, KTH Royal Institute of Technology
and Stockholm University, Roslagstullsbacken 23,
10691 Stockholm, Sweden
}

\date{\today; revised ; accepted  }
\begin{document}

\maketitle

%\preprint{NORDITA-2018-005}

\begin{abstract}
A mean-field theory of differential rotation in a density stratified turbulent convection
has been developed.
This theory is based on a combined effect of the turbulent heat flux and
anisotropy of turbulent convection on the Reynolds stress.
A coupled system of dynamical budget equations consisting in
the equations for the Reynolds stress, the entropy fluctuations
and the turbulent heat flux has been solved.
To close the system of these equations, the spectral tau approach
which is valid for large Reynolds and Peclet numbers, has been applied.
The adopted model of the background turbulent convection takes into account an
increase of the turbulence anisotropy and a decrease of the turbulent
correlation time with the rotation rate.
This theory yields the radial profile of the differential
rotation which is in agreement
with that for the solar differential rotation.
\end{abstract}

\section{Introduction}

Origin of the solar and stellar magnetic fields is associated with
a mean-field dynamo (refereed as $\alpha \Omega$
or $\alpha^2 \Omega$ dynamos) that is based on the combined effect of
helical turbulent motions and a differential rotation
\citep[see, e.g.,][]{moffatt1978,parker1979,krause1980,zeldovich1983,rudiger2013}.
A non-zero mean kinetic helicity produced by a rotating density stratified turbulent
convection, causes the $\alpha$ effect in the solar convective zone.
An origin of the solar differential rotation is
related to an anisotropic eddy viscosity \citep{kippenhahn1963,durney1985,rudiger1980,rudiger1989}.
This idea has been applied in developing a theory of the
differential rotation \citep{durney1993,kichatinov1993,kitchatinov2005}.
The turbulent heat flux in these theories has been introduced
phenomenologically using the mixing-length theory relation:
$\langle {\bf u'}^2 \rangle \propto g \tau_0 \langle u'_z s' \rangle$,
where $\langle u'_z s' \rangle$ is the vertical turbulent heat flux,
${\bf u}'$ and $s'$ are fluctuations of fluid velocity and entropy,
${\bf g}$ is the gravity acceleration  and $\tau_0$ is the characteristic
turbulent time. Also a quasi-linear approach that is valid for small
fluid Reynolds numbers has been applied in these studies.

Additional possibility for the production of the solar differential rotation
is associated with an effect of the turbulent heat flux on the Reynolds
stress in a rotating density stratified turbulent convection.
Based on this idea, \cite{kleeorin2006} develop a mean field theory
of the differential rotation, where a coupled system of dynamical equations
for the Reynolds stress, the entropy fluctuations and the
turbulent heat flux has been solved adopting a spectral $\tau$ approach.
It was demonstrated  \citep{kleeorin2006} that the ratio of the contributions
to the Reynolds stress caused by the turbulent heat flux and the anisotropic
eddy viscosity is of the order of $\sim 10 \, (H_\rho / \ell_0)^2$,
where $\ell_{0}$ is the maximum scale of turbulent motions and
$H_\rho$ is the fluid density variation scale.
This theory allows to determine the profiles of the differential rotation
in the upper part of the solar convection zone where the rotation is slow
in comparison with the turbulent time.

In the low part of the solar convective zone, the rotation is fast
in comparison with the turbulent time.
This causes a strong anisotropy of the turbulent convection
that is an additional source of the solar differential rotation.
One of the key theoretical questions is how can turbulent convection
be modified by the fast rotation, and how can it affect the production
of the differential rotation.
This issue remains to be an open unresolved problem in the solar physics
and astrophysics.
Note that different theories of the solar differential rotation can be validated
using data from the surface measurements of the solar angular velocity
\citep[see, e.g.,][]{howard1970,snodgrass1984} and helioseismology based on measurements
of the frequency of $p$-mode oscillations
\citep[see, e.g.,][]{duvall1986,dziembowski1989,thompson1990,kosovichev1997,schou1998}.

In the present study a combined effect of the
turbulent heat flux and the turbulence anisotropy
increasing with the rotation rate on
the Reynolds stress has been studied for a rotating
density stratified turbulent convection.
The spectral tau approach which is valid for large
Reynolds and Peclet numbers, has been used
in this study.
This allows us to advance the mean-field theory
of the solar differential rotation and obtain
the profiles of the differential rotation versus
radius which are in agreement with the
measured profiles of the solar differential rotation.

\section{Effect of rotation on the Reynolds stress, entropy fluctuations
and turbulent heat flux}

To develop the theory of differential
rotation in a small-scale density stratified turbulent convection,
we use a mean-field approach whereby the
velocity, pressure and entropy are decomposed into
mean and fluctuating parts.
This approach implies that there is a separation of temporal
and spatial scales, so that the mean fields are varied in much larger scales
in comparison with those for fluctuations.

Let us determine the dependencies of the Reynolds stresses
$\langle u'_{i}(t,{\bm x}) \, u'_{j}(t,{\bm x})
\rangle $ and the turbulent heat flux $\langle
s'(t,{\bm x}) \, u'_{i}(t,{\bm x}) \rangle$ on
the mean fields, where angular brackets denote the ensemble
averaging. To this end we use equations for fluctuations of
velocity and entropy in a rotating turbulent
convection, which are obtained by subtracting
equations for the mean fields from the
corresponding equations for the total
fields.
The equations for fluctuations of velocity ${\bm
u}'$ and entropy $s'$ are given by
\begin{eqnarray}
{\partial {\bm u}' \over \partial t} &=& - ({\bm U} \cdot
\bec{\nabla}) {\bm u}' - ({\bm u}' \cdot \bec{\nabla}) {\bm U} -
\bec{\nabla} \biggl({p' \over \rho_{0}}\biggr)
- {\bm g} \, s' + 2 {\bm u}' \times {\bm \Omega} + {\bm U}^{N},
\label{A1} \\
{\partial s' \over \partial t} &=& - {\Omega_{b}^{2} \over g} ({\bm
u}' \cdot {\bm e}) - ({\bm U} \cdot \bec{\nabla}) s' + S^{N} .
\label{A2}
\end{eqnarray}
Equations~(\ref{A1}) and~(\ref{A2}) are written in the reference
frame rotating with the angular velocity ${\bm
\Omega}$. Here $p'$ are fluctuations of fluid
pressure, the entropy fluctuations are
determined by $s' = (\gamma P_{0})^{-1} p' -
\rho_{0}^{-1} \rho'$, the mean fields ${\bm U}$ and $S$
are the mean velocity and entropy, ${\bm
e}$ is the unit vector directed opposite to ${\bm
g}$ and $\Omega_{b}^{2} = - {\bm
g} \cdot \bec{\nabla} S$. The fluid velocity for a low Mach number
flows satisfies the continuity equation written
in the anelastic approximation, ${\rm div} \,
(\rho_0 \, {\bm U}) = 0$ and ${\rm div} \,
(\rho_0 \, {\bm u}') = 0$. The variables with the
subscript $ "0" $ correspond to the hydrostatic
nearly isentropic basic reference state, i.e.,
$\bec{\nabla} P_{0} = \rho_{0} {\bm g}$ and ${\bm
g} \cdot [(\gamma P_{0})^{-1} \bec{\nabla} P_{0}
- \rho_{0}^{-1} \bec{\nabla} \rho_{0}] \approx
0$, where $\gamma$ is the ratio of specific
heats. The turbulent convection is regarded as a
small deviation from a well-mixed adiabatic
reference state.
The nonlinear terms ${\bm U}^{N}$ and $S^{N}$ in
Eqs.~(\ref{A1}) and~(\ref{A2}) which include the molecular
dissipative terms, are given by
\begin{eqnarray*}
U^{N} &=& \langle ({\bm u}' \cdot \bec{\nabla}) {\bm u}'
\rangle - ({\bm u}' \cdot \bec{\nabla}) {\bm u}' + {\bm
f}_{\nu}({\bm u}'),
\\
S^{N} &=& \langle ({\bm u}' \cdot
\bec{\nabla}) s' \rangle - ({\bm u}' \cdot \bec{\nabla}) s' - (1/
T_{0}) \, \bec{\nabla} \cdot {\bm F}_{\kappa}({\bm u}', s'),
\end{eqnarray*}
where $\rho_0 \, {\bm f}_{\nu}({\bm U}) $ is the
mean molecular viscous force, ${\bm
F}_{\kappa}({\bm U}, S) $ is the mean heat flux
associated with the molecular thermal conductivity.

To study the rotating turbulent convection we
perform the derivations which include the
following steps: (i) adopting new variables for
fluctuations of velocity ${\bm v} = \sqrt{\rho_0}
\, {\bm u}' $ and entropy $s = \sqrt{\rho_0} \,
s'$; (ii) derivation of the equations for the
second moments of the velocity
fluctuations $\langle v_i \, v_j \rangle$, the
entropy fluctuations $\langle s^2 \rangle$ and
the turbulent heat flux $\langle v_i \, s
\rangle$ in the ${\bm k}$ space, where we apply a multi-scale approach
\citep{roberts1975}, which separates the mean fields
varied in large scales from fluctuations varied in
small scales; (iii) application of the spectral $\tau$ approximation
and solution of the derived second-moment equations in the
${\bm k}$ space; (iv) returning to the physical
space to obtain formulae for the Reynolds
stress and the turbulent heat flux as the
functions of the rotation rate.

Using Eqs.~(\ref{B1})-(\ref{B2}) for the fluctuations of velocity and entropy
in ${\bm k}$ space derived in Appendix~A, we obtain equations
for the following correlation functions:
$f_{ij}({\bm k},{\bm K}) = \langle v_i(t,{\bm k}_1) v_j(t,{\bm
k}_2) \rangle$,
$F_{i}({\bm k},{\bm K}) = \langle s(t,{\bm k}_1) v_i(t,{\bm
k}_2) \rangle$, and
$\Theta({\bm k},{\bm K}) = \langle s(t,{\bm k}_1) s(t,{\bm
k}_2) \rangle$,
where ${\bm k}_1 = {\bm k} + {\bm K} / 2$ and ${\bm k}_2 = -{\bm k} + {\bm K} / 2$.
Here the wave vectors ${\bm K}$ and ${\bm k}$ are related to the large and small scales, respectively.
Hereafter we omit the argument $t$ in the correlation functions to simplify notations.
The equations for these second moments are given by
\begin{eqnarray}
{\partial f_{ij}({\bm k},{\bm K}) \over \partial t} &=& \left(I_{ijmn}^U +
L_{ijmn}^{\Omega}\right) \, f_{mn} + M_{ij}^F + \hat{\cal N} \tilde f_{ij} ,
\label{B3} \\
{\partial F_{i}({\bm k},{\bm K}) \over \partial t} &=& \left(J_{im}^U +
D_{im}^{\Omega}\right) \, F_{m} + g e_m P_{im}({\bm k}_1) \, \Theta
+ \hat{\cal N} \tilde F_{i} ,
\label{B4} \\
{\partial \Theta({\bm k},{\bm K}) \over \partial t} &=&  - {\rm div} \, \left({\bm U} \Theta \right) + \hat{\cal N} \Theta ,
\label{B5}
\end{eqnarray}
where
\begin{eqnarray*}
I_{ijmn}^U &=& J^U_{im}({\bm k}_1) \, \delta_{jn} + J^U_{jn}({\bm
k}_2) \, \delta_{im}
= \biggl[2 k_{iq} \delta_{mp} \delta_{jn} + 2 k_{jq} \delta_{im}
\delta_{pn} - \delta_{im} \delta_{jq} \delta_{np}
\\
&& \; - \delta_{iq} \delta_{jn} \delta_{mp} + \delta_{im}
\delta_{jn} k_{q} {\partial \over \partial k_{p}} \biggr] \nabla_{p}
U_{q} - \delta_{im} \delta_{jn} \, \left({\rm div} \, {\bm U} + {\bm U}
{\bm \cdot} \bec{\nabla}\right),
\\
M_{ij}^F &=& g e_m [P_{im}({\bm k}_1) F_{j}({\bm
k},{\bm K})  + P_{jm}({\bm k}_2) F_{i}(-{\bm
k},{\bm K})],
\end{eqnarray*}
and $L_{ijmn}^{\Omega} = D_{im}^{\Omega}({\bm k}_1) \, \delta_{jn} +
D_{jn}^{\Omega}({\bm k}_2) \, \delta_{im}$,
$\, J_{ij}^U({\bm k}) = 2 k_{in} \nabla_{j} U_{n} - \nabla_{j} U_{i} -
(1/ 2) \, {\rm div} \, {\bm U} \,  \delta_{ij}$
and $D_{ij}^{\Omega}({\bm k}) = 2 \varepsilon_{ijm} \Omega_n k_{mn}$.
Here $\delta_{ij}$ is the Kronecker tensor, $k_{ij} = k_i  k_j /
k^2$, $\varepsilon_{ijk}$ is the Levi-Civita tensor,
and $F_{i}(-{\bm k},{\bm K}) = \langle s({\bm k}_2) v_i({\bm
k}_1) \rangle$. The correlation functions $f_{ij}$, $\,
F_{i}$ and $\Theta$ are proportional to the non-uniform fluid density
$\rho_0$. Here $\hat{\cal N}\tilde f_{ij}$, $\, \hat{\cal
N}\tilde F_{i}$ and $\hat{\cal N}\Theta$ are the terms which are
related to the third-order moments appearing due to the nonlinear
terms. In particular,
\begin{eqnarray*}
\hat{\cal N}\tilde f_{ij} &=& \langle P_{im}({\bm k}_1)
v^{N}_{m}({\bm k}_1) v_j({\bm k}_2) \rangle
+ \langle v_i({\bm k}_1) P_{jm}({\bm k}_2) v^{N}_{m}({\bm k}_2)
\rangle ,
\\
\hat{\cal N}\tilde F_{i} &=& \langle s^{N}({\bm k}_1) u_j({\bm k}_2)
\rangle + \langle s({\bm k}_1) P_{im}({\bm k}_2) v^{N}_{m}({\bm
k}_2) \rangle ,
\\
\hat{\cal N}\Theta &=& \langle s^{N}({\bm k}_1) s({\bm k}_2) \rangle
+ \langle s({\bm k}_1) s^{N}({\bm k}_2) \rangle .
\end{eqnarray*}

The equations for the second-order moments contain high-order
moments and a closure problem arises \citep[see, e.g.,][]{monin2013,mccomb1990}.
We apply the spectral $\tau$ approximation
that is a sort of third-order closure procedure
\citep[see, e.g.,][]{orszag1970,pouquet1976,kleeorin1990,rogachevskii2004}.
The spectral $\tau$ approximation postulates that the deviations of the
third-order-moment terms, $\hat{\cal N}f_{ij}({\bm k})$, from the
contributions to these terms afforded by the background turbulent
convection, $\hat{\cal N}f_{ij}^{(0)}({\bm k})$, are expressed
through the similar deviations of the second moments, $f_{ij}({\bm
k}) - f_{ij}^{(0)}({\bm k})$, i.e.,
\begin{eqnarray}
\hat{\cal N}f_{ij}({\bm k}) - \hat{\cal N}f_{ij}^{(0)}({\bm k}) = -
{f_{ij}({\bm k}) - f_{ij}^{(0)}({\bm k}) \over \tau_r(k)},
\label{B6}
\end{eqnarray}
and similarly for other tensors, where $\hat{\cal
N}f_{ij} = \hat{\cal N}\tilde f_{ij} +
M_{ij}^F(F^{\Omega=0})$ and $\hat{\cal N} F_{i} =
\hat{\cal N}\tilde F_{i} + g e_n P_{in}(k)
\Theta^{\Omega=0} $, the superscript $ (0) $
corresponds to the background turbulent
convection (i.e., a turbulent convection with $
\nabla_{i} U_{j} = 0)$, $\, \tau_r (k) $ is the
characteristic relaxation time of the statistical
moments, which  can be identified with the
correlation time $\tau(k)$ of the turbulent
velocity field for large Reynolds numbers. The
quantities $F^{\Omega=0}$ and $\Theta^{\Omega=0}$
are for a nonrotating turbulent convection with
nonzero spatial derivatives of the mean velocity.
Validation of the $\tau$ approximation has been done in various numerical
simulations and analytical studies
\citep[see, e.g.,][]{brandenburg2005,brandenburg2004,brandenburg2012,rogachevskii2007,rogachevskii2011,rogachevskii2012,kapyla2012}.
Note that we apply the
$\tau$-approximation~(\ref{B6}) only to study the
deviations from the background turbulent
convection which are caused by the spatial
derivatives of the mean velocity.
The background turbulent convection is assumed to be known (see below).

We use the following model of the background
turbulent convection which  takes into account an
increase of the anisotropy of turbulence with
increase of the rate of rotation:
\begin{eqnarray}
f_{ij}^{(0)} \equiv \langle v_i({\bm k}_1) \,
v_j({\bm k}_2)  \rangle^{(0)} &=& {E(k)
\, [1 + 2 k \, \varepsilon_u \, \delta(k_z)]\over
8 \pi \, k^2 \, (k^2 + \tilde\lambda^2) \, (1 +
\varepsilon_u)} \Big[\delta_{ij} \, (k^2 +
\tilde\lambda^2)  - k_i \, k_j - \tilde\lambda_i
\, \tilde\lambda_j
\nonumber\\
&& + i \, \big(\tilde\lambda_i \,
k_j - \tilde\lambda_j \, k_i\big) \Big] \rho_0 \, \langle {\bm u}'^2 \rangle^{(0)},
\label{B15}\\
F_{i}^{(0)} \equiv \langle v_i({\bm k}_1) \,
s({\bm k}_2)  \rangle^{(0)} &=& {3 \, E(k) \over 8 \pi \,k^4}
\Big[k^2 \, e_j \, P_{ij}({\bm
k}) - i \tilde\lambda  \, k_j \, P_{ij}({\bm e})
\Big] \rho_0 \, \langle s' \, u'_z \rangle^{(0)} ,
\label{B16}
\end{eqnarray}
and $\Theta^{(0)} \equiv \langle s({\bm k}_1) \, s({\bm k}_2) \rangle^{(0)}  =  \rho_0 \, \langle (s')^2 \rangle^{(0)} \, E(k) / 4\pi k^{2}$, where $\bec{\tilde\lambda} =
(\bec{\lambda} - \bec{\nabla}) / 2$, $\,
\bec{\lambda} = -(\bec{\nabla} \rho_0) / \rho_0$.
We assume that the background turbulent convection is the Kolmogorov type turbulence with
a constant flux of energy over the spectrum,
i.e., the kinetic energy spectrum
$E(k) = - d \bar \tau(k) / dk$, $\, \bar \tau(k) =
(k / k_{0})^{1-q}$ with the exponent of the kinetic 
energy spectrum $1 < q < 3$, e.g., $q =5/3$ is for Kolmogorov spectrum.
The turbulent correlation time
$\tau(k) = 2 \tau_{_{\Omega}} \bar \tau(k)$, where
$\tau_{_{\Omega}} = \ell_{0} / u_{0}$, and
$\ell_{0}$ is the energy containing scale of
turbulent motions, $u_{0}=\sqrt{\langle {\bm u}'^2 \rangle^{(0)}}$ is the
characteristic turbulent velocity in the scale
$\ell_{0}$ and $k_{0} = 1 /\ell_{0}$. 
We consider an anisotropic turbulent convection as a combination of a three-dimensional isotropic turbulence and two-dimensional turbulence in the plane perpendicular to the rotational axis. The degree of anisotropy $\varepsilon_u$ is defined as  the ratio of turbulent kinetic energies of two-dimensional to three-dimensional motions.
In this model we neglect effects which are
O$(\lambda^3, \nabla^3 \langle {\bm v}^2 \rangle^{(0)})$.
 
The effect of rotation on the turbulent
correlation time is described just by an heuristic argument, i.e.,
we assume that
$\tau_\Omega^{-2} = \tau_0^{-2} + \Omega^2 /  C_\Omega^{2}$,
that yields:
\begin{eqnarray}
\tau_{_{\Omega}} = {\tau_0 \over \left[1 + \left(C_{_{\Omega}}^{-1} \, \Omega \, \tau_0\right)^2\right]^{1/2}} .
\label{A21}
\end{eqnarray}
This implies that for fast rotation, $\Omega \, \tau_0 \gg 1$, the
parameter $\omega = 8 \Omega  \, \tau_{_{\Omega}}$
tends to be limiting value $\omega_m = 8 C_{_{\Omega}}$, where the dimensionless constant $C_{_{\Omega}} \sim 1$.

The solution of Eqs.~(\ref{B3})--(\ref{B5}) after application of the spectral $\tau$
approximation, and the integration over the ${\bm k}$ space
(see Appendix~\ref{solutions}) allow
us to determine the Reynolds stress and the effective
force versus angular velocity.
The latter yields the mean-field equation for the differential rotation
(see next section), which takes into account the effects of rotating
density stratified turbulent convection.

\section{Mean-field equation for differential rotation}

The differential rotation in the axisymmetric fluid flow is
determined by linearized Navier-Stokes equation for the toroidal component
$U_\varphi(r, \theta) \equiv r \, \sin \theta \, \delta \Omega$ of
the mean velocity:
\begin{eqnarray}
\rho_0 \, {\partial U_\varphi \over \partial t} &=& {1 \over r^3}
{\partial \over \partial r} (r^3 \sigma_{r \varphi}) + {1 \over r
\sin^2 \theta} {\partial \over \partial \theta} (\sin^2 \theta \,
\sigma_{\theta \varphi})
+ 2 \, \rho_0 \, ({\bm U} {\bm \times} {\bm \Omega})_\varphi ,
\label{C1}
\end{eqnarray}
where the tensor $\sigma_{ij} = - \langle v_i \, v_j \rangle$ is
determined by the Reynolds stress:
\begin{eqnarray}
\sigma_{r \varphi} &\equiv& - e_j^\varphi \, e_i^r \, \langle v_i \,
v_j \rangle = \sigma_{r \varphi}^{\nu_T} + \sigma_{r \varphi}^F + \sigma_{r \varphi}^u ,
\label{C2}\\
\sigma_{\theta \varphi} &\equiv& - e_j^\varphi \, e_i^\theta \,
\langle v_i \, v_j \rangle = \sigma_{\theta \varphi}^{\nu_T}
+ \sigma_{\theta \varphi}^F + \sigma_{\theta \varphi}^u .
\label{C3}
\end{eqnarray}
and ${\bm e}^r$,  $\, {\bm e}^\theta$ and $\, {\bm e}^\varphi$ are
the unit vectors along the radial, meridional and toroidal
directions of the spherical coordinates $r, \theta, \varphi$.
There are three contributions to the tensor $\sigma_{ij} = - \langle v_i
\, v_j \rangle$ in Eqs.~(\ref{C2}) and ~(\ref{C3}). The first term in the right hand
side of Eqs.~(\ref{C2}) and ~(\ref{C3}) describes the contribution
$\sigma_{ij}^{\nu_T}$ to the Reynolds stress caused by turbulent viscosity $\nu_{_{T}}$:
\begin{eqnarray}
\sigma_{r \varphi}^{\nu_T} = \rho_0 \, \nu_{_{T}} \, r \, {\partial \over
\partial r} \biggl({U_\varphi \over r}\biggr),
\label{RC2}\\
\sigma_{\theta \varphi}^{\nu_T} = \rho_0 \, \nu_{_{T}} \, {\sin \theta
\over r} \, {\partial \over \partial \theta} \biggl({U_\varphi \over
\sin \theta} \biggr) .
\label{RC3}
\end{eqnarray}
The second term in Eqs.~(\ref{C2}) and ~(\ref{C3}) determines the contribution
$\bec{\sigma}^F$ to the Reynolds stress caused by the turbulent
heat flux:
\begin{eqnarray}
\sigma_{r \varphi}^F &=& {1 \over 6} \, \rho_0 \, \tau_{_{\Omega}}^2 \, g \,
\langle s' \, u'_z \rangle^{(0)} \, \Omega \, \sin \theta \, [\Phi_1(\omega) + \cos^2 \theta
\, \Phi_2(\omega)] \;,
\label{B24}\\
\sigma_{\theta \varphi }^F &=& {1 \over 3} \, \rho_0 \, \tau_{_{\Omega}}^2
\, g \,   \langle s' \, u'_z \rangle^{(0)} \, \Omega \, \sin^2 \theta \, \cos \theta \,
\Phi_2(\omega) ,
\label{B25}
\end{eqnarray}
where the parameter $\omega = 8 \, \Omega \, \tau_{_{\Omega}}$.
The functions $\Phi_1(\omega)$ and $\Phi_2(\omega)$
are given by Eqs.~(\ref{I30})--(\ref{I32}) in Appendix~\ref{solutions} and are shown in Fig.~\ref{Fig1}.
When the turbulent correlation time is independent of the rotation rate,
Eqs.~(\ref{B24}) and~(\ref{B25}) coincide with those obtained by
\cite{kleeorin2006}.

\begin{figure}
\centering
\includegraphics[width=9cm]{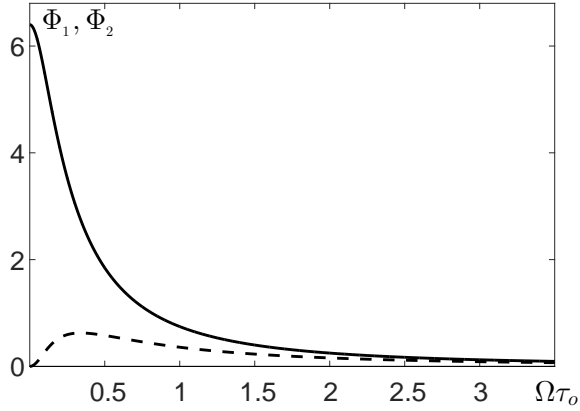}
\caption{\label{Fig1} The functions $\Phi_1(\Omega \tau_0)$ (solid) and $\Phi_2(\Omega \tau_0)$ (dashed) versus $\Omega \tau_0$.}
\end{figure}

The third term in Eqs.~(\ref{C2}) and~(\ref{C3})
determines the contribution $\bec{\sigma}^u$ to the Reynolds
stress caused by the anisotropy of turbulence due to the
nonuniform fluid density and fast uniform rotation (see
Eq.~(\ref{I2}) in Appendix~\ref{solutions}):
\begin{eqnarray}
\sigma_{r \varphi}^u &=& - {\lambda^2 \, \ell_0^2 \over 20} \, \rho_0 \, \langle {\bm u}'^2 \rangle^{(0)} \, \tau_{_{\Omega}} \, \Omega \, \sin \theta \, (1 + \cos^2 \theta) ,
\label{BB24}\\
\sigma_{\theta \varphi }^u &=& {\lambda^2 \, \ell_0^2 \over 20} \, \rho_0 \, \langle {\bm u}'^2 \rangle^{(0)} \, \tau_{_{\Omega}} \, \Omega \, \sin^2 \theta \, \cos \theta .
\label{BB25}
\end{eqnarray}
Equation~(\ref{C1}) in a steady-state that determines the profiles of the differential rotation, reads:
\begin{eqnarray}
\hat {\cal W}(r) \, \rho_0 \nu_{_{T}} \biggl\{{\partial \over \partial r}
{\delta \Omega \over \Omega}
 &+& {1 \over r} \, \Big[a_F \, \left(\Phi_1 (\omega) + \Phi_2 (\omega) X^2 \right)
- 2 a_u \, \lambda^2 \ell_0^2 \left(1 + X^2 \right) \Big] \biggr\}
\nonumber\\
&-& {\rho_0 \nu_{_{T}} \over r^2} \, \hat {\cal M}(X) \, \biggl\{{\delta \Omega \over \Omega} - \Big[a_F \, \Phi_2 (\omega) + a_u \lambda^2 \ell_0^2 \Big] \, X^2 \biggr\} = 0,
\label{C4}
\end{eqnarray}
where the operators $\hat {\cal W}(r)$ and $\hat {\cal M}(X)$ are defined as
\begin{eqnarray*}
\hat {\cal W}(r) f(r)= {1 \over r^4} \, {\partial \over
\partial r} \, \left[r^4 f(r)\right], \quad \quad  \hat {\cal M}(X) \phi(X)= \left[(X^2 - 1) {\partial^2 \over \partial X^2} + 4 X {\partial \over \partial X} \right] \phi(X),
\end{eqnarray*}
$X = \cos\theta$, and the parameters $a_F$ and $a_u$ are given by $a_F = \tau_\Omega^2 \, g \, \langle s' \, u'_z \rangle^{(0)} / 6 \, \nu_{_{T}}$ and
$a_u = \tau_\Omega \, \langle {\bm u}'^2 \rangle^{(0)} / 40 \, \nu_{_{T}}$.
We seek a solution of Eq.~(\ref{C4}) in the form:
\begin{eqnarray}
{\delta \Omega \over \Omega} = \sum_{n=0}^{\infty} \, C_{2n}^{3/2}(X) \,
\tilde \Omega_{2n}(r) ,
\label{C5}
\end{eqnarray}
where the radius $r$ is measured in units of the solar radius $R_\odot$,
and the function $C_n^{3/2}(X)$ satisfies the equation for the
ultra-spherical polynomials:
\begin{eqnarray}
[\hat {\cal M}(X) - n(n+3)] \, C_n^{3/2}(X) = 0  .
\label{C6}
\end{eqnarray}
The function $C_n^{3/2}(X)$ has the following properties:
\begin{eqnarray}
&& \int_{-1}^1 \, (1-X^2) \, C_n^{3/2}(X) \, C_m^{3/2}(X) \,dX
= {(n+1)(n+2) \over n+3/2} \, \delta_{nm} ,
\label{B34}
\end{eqnarray}
$C_0^{3/2}(X) = 1$ and $C_2^{3/2}(X) = (3/2) (5X^2 - 1)$.
Substituting Eq.~(\ref{C5}) into Eq.~(\ref{C4}), we obtain equations for
the functions $\tilde \Omega_0(r)$:
\begin{eqnarray}
&& \tilde \Omega_0(r) = \tilde \Omega_\ast - {1 \over 5}\int^{1}_{r/R_\odot} \Big\{12 a_u \lambda^2 \ell_0^2 - a_F \,\left[5\Phi_1 (\omega) + \Phi_2 (\omega)\right] \Big\} \, {dr \over r},
\label{C10}
\end{eqnarray}
and $\tilde \Omega_2(r)$:
\begin{eqnarray}
&& \hat {\cal W}(r) \, \rho_0 \nu_{_{T}} \, \biggl[{\partial \tilde \Omega_2(r) \over \partial r} + {2 \over 15 \, r} \, \Big(a_F \, \Phi_2 (\omega) - 2 a_u \lambda^2 \ell_0^2 \Big)\biggr]
\nonumber\\
&& \quad \quad \quad \quad - {10 \, \rho_0 \nu_{_{T}} \over r^2}  \biggl[\tilde \Omega_2(r) - {2 \over 15} \left(a_F \, \Phi_2 (\omega) + a_u \lambda^2 \ell_0^2\right)\biggr]
= 0 ,
\label{C11}
\end{eqnarray}
where $\tilde \Omega_\ast$ is the free constant determined by the surface boundary condition.

\begin{figure}
\centering
\includegraphics[width=9cm]{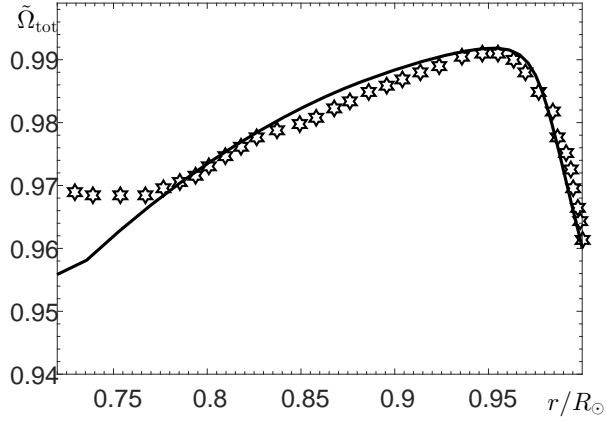}
\caption{\label{Fig2} The total angular velocity $\tilde \Omega_{\rm tot} = \tilde \Omega_0 + 1$ that includes the uniform rotation $\Omega$ versus the radius $r/R_\odot$ (solid). This theoretical profile is compared with the radial profile of the solar angular velocity
obtained from the helioseismology observational data (stars) at the latitude $\phi=30^\circ$ and normalized by the solar rotation frequency $\Omega_\odot(\phi=0)$ at the equator, where $R_\odot$ is the solar radius.}
\end{figure}

In Fig.~\ref{Fig2} we show the total angular velocity $\tilde \Omega_{\rm tot} = \tilde \Omega_0 + 1$ that includes the uniform rotation $\Omega$ versus the radius $r/R_\odot$. This theoretical profile is compared with the radial profile of the solar angular velocity
obtained from the helioseismology observational data \citep{kosovichev1997} specified for the latitude $\phi=30^\circ$ and normalized by the solar angular velocity $\Omega_\odot(\phi=0)$ at the equator. Note that at $\phi=30^\circ$ the contribution from the term $C_{2}^{3/2}(X) \,
\tilde \Omega_{2}(r)$ to the differential rotation vanishes, because the function $C_2^{3/2}(X) = (3/2) (5X^2 - 1)$ at the angle around $\phi=30^\circ$ vanishes.
To determine $\tilde \Omega_{\rm tot}$ we use the rotation rate dependence of the
turbulent viscosity $\nu_{_{T}}(\omega)=\nu_{_{T}}^{\ast} \Phi_\nu(\omega)$, where
$\nu_{_{T}}^{\ast} = \tau_0 \langle {\bm u}'^2 \rangle^{(0)} / 6$, the functions $\Phi_\nu(\omega)$ is given by Eq.~(\ref{I33}) in Appendix~\ref{solutions} and is shown in Fig.~\ref{Fig3}. Strong change of the turbulent viscosity is caused by the fast rotation during the transition from isotropic three-dimensional turbulence to strongly anisotropic quasi two-dimensional turbulence.

\begin{figure}
\centering
\includegraphics[width=9cm]{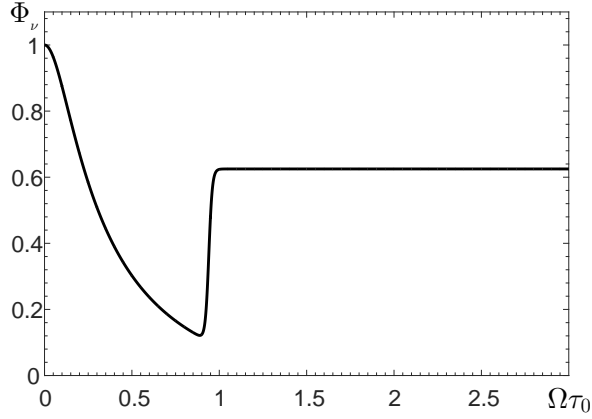}
\caption{\label{Fig3} The rotation rate dependence of the
functions $\Phi_\nu(\Omega \tau_0)$, where $\nu_{_{T}}(\Omega \tau_0)=\nu_{_{T}}^{\ast} \Phi_\nu(\Omega \tau_0)$.}
\end{figure}

\begin{figure}
\centering
\includegraphics[width=9cm]{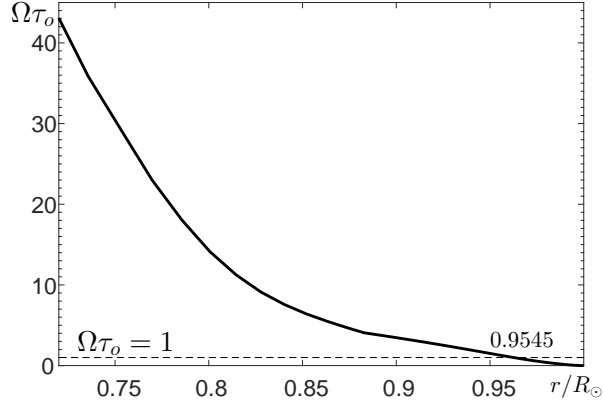}
\caption{\label{Fig4} The profile of $\Omega \tau_0$ versus $r/R_\odot$ based on the
model of the solar convective zone by \cite{spruit1974}.}
\end{figure}

\begin{figure}
\centering
\includegraphics[width=9cm]{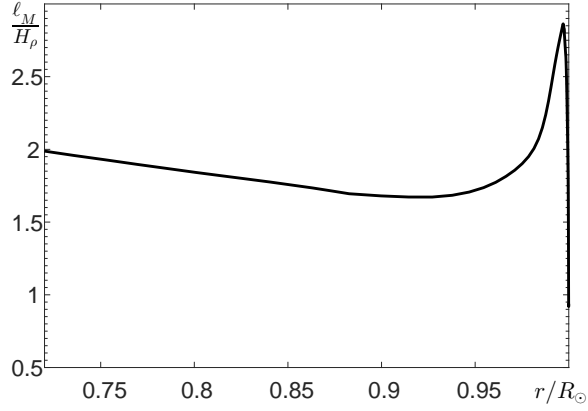}
\caption{\label{Fig5} The profile of the ratio  $\ell_M / H_\rho$ of the mixing length $\ell_M$ to the density stratification length $H_\rho$ versus $r/R_\odot$ that is based on the
model of the solar convective zone by \cite{spruit1974}.}
\end{figure}

For the comparison of the theoretical profiles of the differential rotation and observational data, we use the radial profiles of $\Omega \, \tau_0(r)$ (see Fig.~\ref{Fig4}) and the ratio  $\ell_M(r) / H_\rho(r)$ (see Fig.~\ref{Fig5}) of the mixing length $\ell_M$ to the density stratification length $H_\rho$ based on the model of the solar convective zone by \cite{spruit1974}.
Inspection of Fig.~\ref{Fig2} demonstrates that the theoretical profile of the differential
rotation is in agreement with the profile of the solar differential rotation when
$\ell_M /\ell_0 = 5$. The latter is justified by the results of analytical study \citep{elperin2002,elperin2006} and laboratory experiments \citep{bukai2009}, which show that the integral scale $\ell_0$ of the turbulent convection is smaller in 5 times in comparison with the size of the coherent structures (the large-scale circulations).
We compare the theoretical and observation profiles of the differential rotation for the latitude $\phi=30^\circ$ because for the latitudes which are far from $\phi=30^\circ$, the contribution of the term $\propto \tilde \Omega_2$ (determined by Eq.~(\ref{C11})) to the differential rotation cannot be ignored. More detail comparison of the theoretical and observation profiles of the differential rotation for different latitudes requires the mean-field numerical modelling that is a subject of the separate study.

\section{Conclusions}

We discuss a new theory of differential rotation
based on a combined effect of the turbulent heat flux and
the turbulence anisotropy increasing with the rate of rotation on the Reynolds stress
in a density stratified turbulent convection.
We solve a coupled system of dynamical budget equations which includes
the equations for the Reynolds stress, the entropy fluctuations
and the turbulent heat flux, applying
a spectral $\tau$ approach to close the system of these equations.
The model of the background turbulent convection takes into account an
increase of the turbulence anisotropy and a decrease of the turbulent
correlation time with the rotation rate.
This theory allows to obtain the profile of the differential
rotation versus radius which is in agreement with the
profile of the solar differential rotation.

The mechanism of the differential rotation that is related to
the effect of the turbulent heat flux on Reynolds stress in a
rotating turbulent convection is as follows. The total angular velocity
includes the uniform rotation $\Omega$ and the
differential rotation $\delta \Omega$.
The uniform rotation results in the counter-rotation turbulent
heat flux $\langle s' \, u'_\varphi \rangle $ that is
directed opposite to the uniform rotation $\Omega$.
The counter-rotation  turbulent heat flux is similar to the
counter-wind turbulent heat flux that is
directed opposite to the mean wind known in the atmospheric physics  \citep{elperin2002,elperin2006}.
In turbulent convection an ascending fluid element obeys
larger temperature than the temperature of the surrounding fluid and smaller
toroidal fluid velocity, while a descending fluid element obeys
smaller temperature and larger toroidal fluid velocity. This results in
the turbulent heat flux in the direction opposite to the uniform rotation.
The entropy fluctuations produce fluctuations of the buoyancy force,
that increases fluctuations of the vertical and
meridional components of the velocity which are correlated with the
fluctuations of the toroidal component of the velocity. This implies that
the off-diagonal components of the Reynolds stress,
$\langle u'_r u'_\varphi \rangle$ and $\langle u'_\theta u'_\varphi
\rangle $ are non-zero, producing the toroidal component of the effective force.
The latter results in the formation of the differential rotation
$\delta \Omega$ in turbulent convection.

\begin{acknowledgements}
This work was supported in part by the Research Council of Norway
under the FRINATEK (grant No. 231444).
The authors acknowledge the hospitality of NORDITA and Ural Federal University.
\end{acknowledgements}

\appendix

\section{Derivation of equations for the second moments}

Equations~(\ref{A1}) and (\ref{A2}) in the new variables for
fluctuations of velocity ${\bm v}= \sqrt{\rho_0}
\, {\bm u}' $ and entropy $s = \sqrt{\rho_0} \,
s'$ are given by
\begin{eqnarray}
{1 \over \sqrt{\rho_0}} {\partial {\bm v}({\bm x},t) \over
\partial t} &=& - \bec{\nabla} \biggl({p' \over \rho_0}\biggr)
+ {1 \over \sqrt{\rho_0}} \biggl[2 {\bm v} {\bm \times} {\bm \Omega} -
({\bm v} \cdot \bec{\nabla}) {\bm U}
- G^U \, {\bm v} - {\bm g} \, s \biggr] + {\bm F}_M + {\bm v}^N,
\nonumber\\
\label{M1} \\
{\partial s({\bm x},t) \over \partial t} &=& - {\Omega_{b}^{2} \over
g} ({\bm v} \cdot {\bm e}) - G^U \, s + s^N,
\label{M2}
\end{eqnarray}
where $G^U = (1 / 2) \, {\rm div} \, {\bm U} + {\bm U} {\bm \cdot}
\bec{\nabla}$, $\,  {\bm v}^{N} $ and $ s^{N} $ are the nonlinear
terms which include the molecular viscous and dissipative terms. The fluid velocity
fluctuations ${\bm v}$ satisfy the equation $\bec{\nabla} \cdot {\bm
v} = {\bm v} \cdot {\bm \lambda}/2$.

Let us derive equations for the second-order moments. For this
purpose we rewrite the momentum equation and the entropy equation in
a Fourier space. In particular,
\begin{eqnarray}
{dv_i({\bm k}) \over dt} &=& [D_{im}^{\Omega}({\bm k}) +
\tilde J_{im}^U({\bm k})] v_m({\bm k}) + g \, e_m \, P_{im}({\bm k}) \,
s({\bm k}) + v_i^N({\bm k}),
\label{B1}\\
{ds({\bm k}) \over dt} &=& - G^U({\bm k}) \, s({\bm k}) + s^N ,
\label{B2}
\end{eqnarray}
where
\begin{eqnarray*}
\tilde J_{ij}^U({\bm k}) = 2 k_{in} \nabla_{j} U_{n} - \nabla_{j} U_{i} -
G^U({\bm k}) \delta_{ij},
\quad G^U({\bm k}) = {1 \over 2} \, {\rm div} \, {\bm U} + i ({\bm U}
{\bm \cdot} {\bm k}),
\end{eqnarray*}
$D_{ij}^{\Omega}({\bm k}) = 2 \varepsilon_{ijm} \Omega_n k_{mn}$,
$ \delta_{ij} $ is the Kronecker tensor, $k_{ij} = k_i  k_j /
k^2$ and $\varepsilon_{ijk}$ is the Levi-Civita tensor.
To derive Eq.~(\ref{B1}) we multiply the momentum equation written
in ${\bm k}$-space by $ P_{ij}({\bm k}) = \delta_{ij} - k_{ij} $
to exclude the pressure term.
We also use the following identities:
\begin{eqnarray*}
&& \sqrt{\rho_0} \, \big[\bec{\nabla} {\bm \times} [\bec{\nabla}
{\bm \times} ({\bm u}' {\bm \times} {\bm \Omega})] \big]
= \big({\bm \Omega}  {\bm \times} \bec{\nabla}^{(\lambda)}\big)
\, (\bec{\lambda} {\bm \cdot} {\bm v})
+ \big({\bm \Omega} {\bm \cdot} \bec{\nabla}^{(\lambda)}\big) \,
\big(\bec{\nabla}^{(\lambda)}  {\bm \times} {\bm v}\big),
\\
&& \sqrt{\rho_0} \, \big[\bec{\nabla} {\bm
\times} [\bec{\nabla} {\bm \times} ({\bm g}
\,s')] \big] = - g_j \, \big[ \delta_{ij} \,
\big(\bec{\nabla}^{(\lambda)}\big)^2 -
\nabla_i^{(\lambda)} \nabla_j^{(\lambda)} \big] s,
\\
&& \sqrt{\rho_0} \, \big[\bec{\nabla} {\bm
\times} [\bec{\nabla}  {\bm \times} ({\bm u})]
\big]_{\bm k} = - \big[\Lambda^2 \, \delta_{ij} -
\Lambda_i \lambda_j \big] v_j({\bm k}),
\end{eqnarray*}
where $\bec{\nabla}^{(\lambda)} = \bec{\nabla}  +
\bec{\lambda} /2$, $\, \bec{\lambda} =
-(\bec{\nabla} \rho_0) / \rho_0$, $\,
\bec{\Lambda} = i {\bm k} + \bec{\lambda} /2$.
Using Eqs.~(\ref{B1}) and~(\ref{B2}) we derive equations for the second moments
which are given by Eqs.~(\ref{B3})--(\ref{B5}).

\section{Solutions for the second moments}
\label{solutions}

Equations~(\ref{B3})-(\ref{B5}) in a
steady state and after  applying the spectral
$\tau$ approximation~(\ref{B6}), read
\begin{eqnarray}
f_{ij}({\bm k}) &=& L_{ijmn}^{-1}
\big[f^{(0)}_{mn} + \tau \,  \tilde M_{mn}^F +
\tau \, (I_{mnpq}^U + L_{mnpq}^{\nabla}
+ L_{mnpq}^{\lambda}  + L_{mnpq}^{\nabla^2} +
L_{mnpq}^{\lambda^2})  \, f_{pq} \big],
\nonumber\\
\label{B7}\\
F_i({\bm k}) &=& D_{im}^{-1}
\big[F^{(0)}_{m}({\bm k})  + \tau \, (J_{mn}^U +
D_{mn}^{\nabla} + D_{mn}^{\lambda}
+ D_{mn}^{\nabla^2} + D_{mn}^{\lambda^2}) F_{n} \big],
\label{B8}
\end{eqnarray}
and $\Theta({\bm k}) = [1 - \tau ({\bm U} {\bm \cdot} \bec{\nabla})]
\Theta^{(0)}({\bm k})$,
where
\begin{eqnarray}
\tilde M_{ij}^F &=& g e_m \big\{\big[P_{im}({\bm
k}) + k_{im}^\nabla  + k_{im}^\lambda -
k_{im}^{\nabla^2} + k_{im}^{\lambda^2} \big]
\tilde F_{j}({\bm k})
+ \big[P_{jm}({\bm k}) - k_{jm}^\nabla
\nonumber\\
&& -k_{jm}^\lambda - k_{jm}^{\nabla^2}  + k_{jm}^{\lambda^2} \big] \tilde F_{i}(-{\bm k})
\big\},
\label{BB8}
\end{eqnarray}
and $\tilde F_{i}=F_{i}-F_{i}^{\Omega=0}$ and we
neglected  terms $\sim O(\nabla^3, \lambda^3)$.
Here the operator $D_{ij}^{-1}= \chi(\psi) \, (\delta_{ij} +
\psi \, \varepsilon_{ijm} \, \hat k_m + \psi^2 \, k_{ij})$ is the inverse of
$\delta_{ij} - \tau \tilde D_{ij}$ and the
operator $L_{ijmn}^{-1}({\bm \Omega})$ is the
inverse of $\delta_{im} \delta_{jn} - \tau \,
\tilde L_{ijmn}$ \citep{kleeorin2003,elperin2005},
where
\begin{eqnarray}
L_{ijmn}^{-1}({\bm \Omega}) &=&  {1 \over 2} [B_1
\, \delta_{im} \delta_{jn} + B_2 \, k_{ijmn} +
B_3 \, (\varepsilon_{imp} \delta_{jn}
+ \varepsilon_{jnp} \delta_{im}) \hat k_p +B_4 \, (\delta_{im} k_{jn}
\nonumber\\
&& + \delta_{jn} k_{im}) + B_5 \, \varepsilon_{ipm} \varepsilon_{jqn}
k_{pq} + B_6 \, (\varepsilon_{imp} k_{jpn} +
\varepsilon_{jnp} k_{ipm}) ] ,
\label{B14}
\end{eqnarray}
and $\hat k_i = k_i / k$, $\, \chi(\psi) = 1 / (1
+ \psi^2) $, $\, \psi = 2 \tau(k) \, ({\bm k}
\cdot {\bm \Omega}) / k $, $\, B_1 = 1 + \chi(2
\psi) ,$ $\, B_2 = B_1 + 2 - 4 \chi(\psi) ,$ $\,
B_3 = 2 \psi \, \chi(2 \psi) ,$ $\, B_4 = 2
\chi(\psi) - B_1 ,$ $\, B_5 = 2 - B_1 $ and $B_6
= 2 \psi \, [\chi(\psi) - \chi(2 \psi)]$.

To obtain solutions for the second moments,
we extract in tensors $D_{ij}^{\Omega}$ and
$L_{ijmn}^{\Omega}$  the parts which
depend on large-scale spatial derivatives
and on the density stratification effects:
\begin{eqnarray}
D_{ij}^{\Omega} &=& \tilde D_{ij} + D_{ij}^\nabla
+  D_{ij}^{\nabla^2} + D_{ij}^\lambda +
D_{ij}^{\lambda^2} + O(\nabla^3) ,
\label{B18}\\
L_{ijmn}^{\Omega} &=& \tilde L_{ijmn}+
L_{ijmn}^\nabla  + L_{ijmn}^{\nabla^2} +
L_{ijmn}^\lambda + L_{ijmn}^{\lambda^2}
+ O(\nabla^3) ,
\label{B19}
\end{eqnarray}
where
\begin{eqnarray*}
\tilde L_{ijmn} &=& 2 \, \Omega_q \,
(\varepsilon_{imp}  \, \delta_{jn} +
\varepsilon_{jnp} \, \delta_{im}) \, k_{pq},
\quad
L_{ijmn}^\nabla = - 2\,\Omega_q \,
(\varepsilon_{imp}  \, \delta_{jn} -
\varepsilon_{jnp} \, \delta_{im}) \,
k_{pq}^\nabla ,
\\
L_{ijmn}^\lambda &=& - 2\,\Omega_q \,
\Big[(\varepsilon_{imp}  \, \delta_{jn} -
\varepsilon_{jnp} \, \delta_{im}) \,
k_{pq}^\lambda
+{i \over k^2} (\varepsilon_{ilq} \,
\delta_{jn}   \,\lambda_m- \varepsilon_{jlq} \,
\delta_{im}\,\lambda_n) \, k_{l} \Big],
\\
L_{ijmn}^{\nabla^2} &=& 2\, \Omega_q \,
(\varepsilon_{imp}  \, \delta_{jn} +
\varepsilon_{jnp} \, \delta_{im}) \,
k_{pq}^{\nabla^2},
\quad
L_{ijmn}^{\lambda^2} = 2\, \Omega_q \,
(\varepsilon_{imp}  \, \delta_{jn} +
\varepsilon_{jnp} \, \delta_{im}) \,
k_{pq}^{\lambda^2},
\end{eqnarray*}
and $\tilde D_{ij} = 2 \varepsilon_{ijp} \Omega_q k_{pq}$,
$D_{ij}^\nabla = 2 \varepsilon_{ijp} \Omega_q k_{pq}^\nabla$,
$D_{ij}^\lambda = 2 \varepsilon_{ijp} \Omega_q k_{pq}^\lambda$,
$D_{ij}^{\nabla^2} = 2 \varepsilon_{ijp} \Omega_q k_{pq}^{\nabla^2}$,
and $D_{ij}^{\lambda^2} = 2 \varepsilon_{ijp} \Omega_q k_{pq}^{\lambda^2}$.
Here
\begin{eqnarray*}
k_{ij}^\nabla &=& {i \over 2 k^2} \, [k_i \nabla_j
+ k_j \nabla_i - 2 k_{ij} ({\bm k} {\bm \cdot} \bec{\nabla})],
\quad
k_{ij}^\lambda = {i \over 2 k^2} \, [k_i \lambda_j
+ k_j \lambda_i - 2 k_{ij} ({\bm k} {\bm \cdot} \bec{\lambda})],
\\
k_{ij}^{\nabla^2} &=& {1 \over 4 k^2} \,
\big[k_{ij} \bec{\nabla}^2  + 2 (k_{ip} \nabla_j
+ k_{jp} \nabla_i) \, \nabla_p
- 4 k_{ijpq} \nabla_p \nabla_q - \nabla_i \nabla_j \big],
\\
k_{ij}^{\lambda^2} &=& {1 \over 4 k^2}
\big[\lambda_i \tilde \nabla_j  + \lambda_j
\tilde \nabla_i - 2 \lambda_m (k_{im} \tilde
\nabla_j + k_{jm} \tilde \nabla_i
+ k_{ij} \tilde\nabla_m) + \lambda_i \lambda_j -
k_{ij} \lambda^2  + 4 k_{ijpq}\lambda_p
\lambda_q\big],
\end{eqnarray*}
and $\tilde \nabla_i = \nabla_i - 4 k_{il} \nabla_l$.

After integration in ${\bm k}$ space we obtain
contributions to the Reynolds stress caused by
turbulence anisotropy due to the rapid rotation:
\begin{eqnarray}
f_{ij}^{(u)} &=& {\lambda \over 20} \biggl\{
\left[\left(\hat{\bm \omega} {\bm \times} {\bm e}\right)_i e_j +
\left(\hat{\bm \omega} {\bm \times} {\bm e}\right)_j e_i \right]
\, (\lambda- \nabla_z)
+ \left(\hat{\bm \omega} {\bm \cdot} {\bm e}\right)
\Big[\left(\hat{\bm \omega} {\bm \times} {\bm e}\right)_i
\hat{\omega}_j
\nonumber\\
&& + \left(\hat{\bm \omega} {\bm \times} {\bm e}\right)_j
\hat{\omega}_i \Big] \, (\lambda+ \nabla_z)
\biggr\} {\varepsilon_u \over 1+ \varepsilon_u} \,
\rho_0 \, \langle {\bm u}'^2 \rangle^{(0)} \, \Omega \tau_{_{\Omega}} \, \ell_0^2 .
\label{I2}
\end{eqnarray}
To derive Eq.~(\ref{I2}), we use the following integrals:
\begin{eqnarray*}
&& \int k_{ij}^{\perp} \,d\varphi = \pi \delta_{ij}^{(2)},
\quad \int k_{ijmn}^{\perp} \,d\varphi = {\pi \over 4} \Delta_{ijmn}^{(2)},
\end{eqnarray*}
where $\delta_{ij}^{(2)} \equiv P_{ij}(\Omega) = \delta_{ij} - \Omega_i \Omega_j /\Omega^2$
and $\Delta_{ijmn}^{(2)} = \delta_{ij}^{(2)}\delta_{mn}^{(2)}
+ \delta_{im}^{(2)} \delta_{jn}^{(2)}+ \delta_{in}^{(2)} \delta_{jm}^{(2)}$.

The contributions to the Reynolds stress caused by
the turbulent heat flux are given by Eqs.~(\ref{B24}) and~(\ref{B25}),
where the functions $\Phi_1(\omega)$ and $\Phi_2(\omega)$ are given by
\begin{eqnarray}
\Phi_1(\omega) &=& 2 \Psi_1(\omega) + \Psi_2(\omega/2), \quad
\Phi_2(\omega) = 2 \Psi_2(\omega) + \Psi_2(\omega/2),
\label{I30}\\
\Psi_1(\omega) &=& - {6 \over \omega^{4}}
\biggl[{\arctan \, \omega \over \omega} (1 + \omega^{2}) - {8
\omega^{2}\over 3} -1
+ 2 \, \omega \, Y(\omega) \biggr],
\label{I31}\\
\Psi_2(\omega) &=& {6 \over \omega^{4}} \biggl[5 \, {\arctan \,
\omega \over \omega} (1 + \omega^{2}) + {8 \omega^{2}\over 3} - 5
- 6 \, \omega \, Y(\omega) \biggr] ,
\label{I32}
\end{eqnarray}
$\omega = 8 \Omega \tau_{_{\Omega}}$ and $Y(\omega) = \int_{0}^{\omega}
[\arctan \, y / y] \,d y$.
When the turbulent correlation time is independent of the rotation rate,
Eqs.~(\ref{B24}) and~(\ref{B25}) coincide with those obtained by \cite{kleeorin2006}.

To determine the profile of the differential rotation, we use
the rotation rate dependence of the turbulent viscosity $\nu_{_{T}}(\omega)
=\nu_{_{T}}^{\ast} \Phi_\nu(\omega)$, where $\nu_{_{T}}^{\ast} = \tau_0 \langle {\bm
u}'^2 \rangle^{(0)} / 6$ and the functions $\Phi_\nu(\omega)$ is given by
\begin{eqnarray}
\Phi_\nu(\omega) &=& {1 \over 8(1+ \varepsilon_u)} \, \Big\{(q+3) \, \varepsilon_u + 2 \, \left[A_{1}^{(1)}(\omega) - A_{1}^{(1)}(0) + (q+ 2) C_{1}^{(1)}(0)
+ C_{1}^{(1)}(\omega)\right]
\nonumber\\
&& + A_{2}^{(1)}(\omega) + C_{3}^{(1)}(\omega)\Big\} .
\label{I33}
\end{eqnarray}
Here
\begin{eqnarray*}
A_{1}^{(1)}(\omega) &=& 12 \biggl[{\arctan (\omega) \over \omega}
\biggl(1 - {1 \over \omega^{2}} \biggr)
+ {1 \over \omega^{2}}\left[1 - \ln\left(1 + \omega^{2}\right)\right] \biggr] ,
\\
A_{2}^{(1)}(\omega) &=& - 12 \biggl[{\arctan (\omega) \over
\omega} \biggl(1 - {3 \over \omega^{2}} \biggr)
+ {1 \over \omega^{2}}\left[3 - 2 \ln\left(1 + \omega^{2}\right)\right] \biggr] \;,
\\
C_{1}^{(1)}(\omega) &=& {\arctan (\omega) \over \omega} \biggl(3 -
{6 \over \omega^{2}} - {1 \over \omega^{4}} \biggr)
+ {1 \over \omega^{2}} \biggl({17 \over 3} + {1 \over
\omega^{2}} - 4 \ln\left(1 + \omega^{2}\right) \biggr) ,
\end{eqnarray*}
where $C_{3}^{(1)}(\omega) = A_{1}^{(1)}(\omega) - 5 C_{1}^{(1)}(\omega)$,
and we use equations derived by \cite{elperin2005}, which are adopted for the spherical
geometry.

%\newpage

\bibliographystyle{jpp}

\bibliography{New-dif-rot-JPP}

\end{document}